\documentstyle[epsf,twocolumn,eqsecnum,prb,aps]{revtex}

\begin{document}

\draft

\title{
Vortex structure in chiral $p$-wave superconductors
}

\author{Mitsuaki Takigawa, Masanori Ichioka, Kazushige Machida}
\address{Department of Physics, Okayama University,
         Okayama 700-8530, Japan}
\author{Manfred Sigrist}
\address{Theoretische Physik, Eidgen$\ddot{o}$ssische Technische Hochschule,
 CH-8093 Z$\ddot{u}$rich, Switzerland}

\date{\today}

\maketitle

\begin{abstract}
We investigate the vortex structure in chiral $p$-wave superconductors
by the Bogoliubov-de Gennes theory on a tight-binding model.
We calculate the spatial structure of the pair potential
and electronic state around a vortex, including the anisotropy of
the Fermi surface and superconducting gap structure.
The differences of the vortex structure between
$\sin p_x + {\rm i} \sin p_y$-wave and $ \sin p_x - {\rm i} \sin p_y$-wave
superconductors are clarified in the vortex lattice state.
We also discuss the winding $\mp 3$ case of the
$\sin{(p_x+p_y)} \pm {\rm i} \sin{(-p_x+p_y)}$-wave superconductivity.

\end{abstract}

\pacs{PACS numbers: 74.60.Ec, 74.20.Rp, 74.70.Pq, 74.25.Jb}

\narrowtext

%
%
%
%
%
%
%
%

\section{Introduction}
\label{sec:introduction}

Recently, much attention has been focused on the superconductivity
in quasi-two-dimensional metal
${\rm Sr_2RuO_4}$,~\cite{Maeno} there is experimental evidence that
it might be a realization of a chiral $p$-wave superconductor.\cite{Rice}
The ${\rm ^{17}O}$-NMR measurements reported that there is no reduction
of the Knight shift in the superconducting state,
supporting the spin triplet pairing.~\cite{Ishida}
Moreover $\mu$SR measurements claim that spontaneous magnetic moment
appear in the superconducting state, suggesting a pairing state
with broken time reversal symmetry.~\cite{Luke}
Therefore, the symmetry of superconductivity is likely to be that of
the chiral $p$-wave with the basic form $\Delta_{\pm}\sim p_x \pm{\rm
  i} p_y$ and inplane equal-spin pairing. This simplified form shall,
however, not preclude any details of the
momentum dependence of the quasiparticle gap.
Detailed discussions on the pairing symmetry were given in Refs.
\onlinecite{Machida,Sigrist,Hasegawa,Graf,Zhitomirsky}.

At zero field, $\Delta_{+}$ and $\Delta_{-}$ state are degenerate and,
in general, domain formation of the two states, $\Delta_{+}$ and
$\Delta_{-}$, is expected.
This degeneracy is removed under an external magnetic field with a
component perpendicular to the basal plane,
since $\Delta_{\pm}$ corresponds to states with an orbital angular
momentum along the $z$-axis. Also the vortex structure in the mixed phase
is different for $\Delta_{+}$ and $\Delta_{-}$ and we may consider the
so-called positive vortex (P-vortex)
and negative vortex (N-vortex).~\cite{MatsumotoHeeb}
In the former (latter) corresponds to a vortex with winding
orientation parallel (antiparallel) to the internal winding of the
Cooper pair.
If the external field ${\bf H} $ is parallel to $ \hat{z}$ for $e>0$
($2e$ is the charge of the Cooper pair) or, if ${\bf H} $ is
antiparallel to $ \hat{z}$ for $e<0$, the vortex with a winding
$+1$ appears, i.e.,  $\Delta({\bf r})\sim f(r){\rm e}^{{\rm i} \phi}$
where ${\bf r}=(r \cos\phi , r\sin \phi)$.
This corresponds to the P-vortex (N-vortex) for the phase $\Delta_{+}$
($\Delta_{-}$). For reversed magnetic field direction, one finds
vortices with opposite winding $-1$, thus $\Delta({\bf r}) \sim
f(r){\rm e}^{-{\rm i} \phi}$, which represents the
N-vortex (P-vortex) for $\Delta_{+}$ ($\Delta_{-}$).

The differences of the structure between P-vortex and N-vortex
were studied within the Ginzburg-Laudau (GL) theory~\cite{Heeb,HeebD}
and the quasiclassical theory.~\cite{Kato,KatoHayashi}
Furthermore the electronic state around the vortex core was studied
by the Bogoliubov-de Gennes (BdG) theory in the single vortex case
and continuum model.~\cite{MatsumotoHeeb,HeebD,Matsumoto}
It was suggested that the character of the quantized energy level
for bound state of quasiparticles around the vortex core
is different for P-vortex and N-vortex.

In this paper we investigate the difference of the
vortex structure between the P-vortex and N-vortex, based on the BdG theory
for a tight-binding model,
considering the form of pair potential and electronic states
in the vortex lattice state.
So far, this method has been used to the study of the $d_{x^2-y^2}$-wave
superconductivity in the high-$T_c$ superconductors, applied to the
extended Hubbard model~\cite{Wang,Takigawa,TakigawaJPSJ}
or the $t$-$J$ model.~\cite{Himeda,Ogata}
Here, we introduce the spin triplet $p$-wave pairing interaction instead
of the singlet $d$-wave and $s$-wave pairings.

Using the tight-binding model allows us to take certain aspects of the
real band structure into account. While Sr$_2$RuO$_4$ is a metal with
three electronic bands, we will restrict to one band only. This is
justified from the point of view that one of the three bands dominates
the superconducting properties.\cite{Zhitomirsky,ODS} We take the nearly cylindrical
symmetric band originating from the Ru$^{4+}$-$4d_{t_{2g}}$ orbital
$d_{xy} $ yielding the so-called $ \gamma $-band.
\cite{Mackenzie,MackenzieJS,Puchkov}
The square lattice tight-binding model includes
nearest-neighbor (NN) and next-nearest-neighbor (NNN) hopping, using the
parametrization given, for example, in Ref. \onlinecite{Ng}.
Any dispersion along the $z$-axis is neglected.
The pairing interaction is taken on the
lattice as real space attractive NN interaction yielding the gap
function form $\sin p_x \pm {\rm i} \sin p_y$.\cite{Ng,Miyake}
The superconducting gap is nodeless on the Fermi surface,
but has a strong anisotropy.

Our self-consistent calculation of the vortex structure reveals the
anisotropy of the gap and the band structure, when we consider the
local density of states (LDOS) around the vortex core. Moreover, we
observe the detailed structure of the pair potential in the vicinity
of the vortex core, which contains induced components of $\Delta_{-}$
($\Delta_{+}$) in the background of dominant $\Delta_{+}$
($\Delta_{-}$) superconductivity. Since our discussion includes the
complete vortex lattice, we are able to discuss the
contribution of the inter-vortex transfer of low-energy
quasiparticles bound at the vortex core.

\begin{figure}[tbc]
\begin{center}
\leavevmode
\epsfxsize=5.5cm
\epsfbox{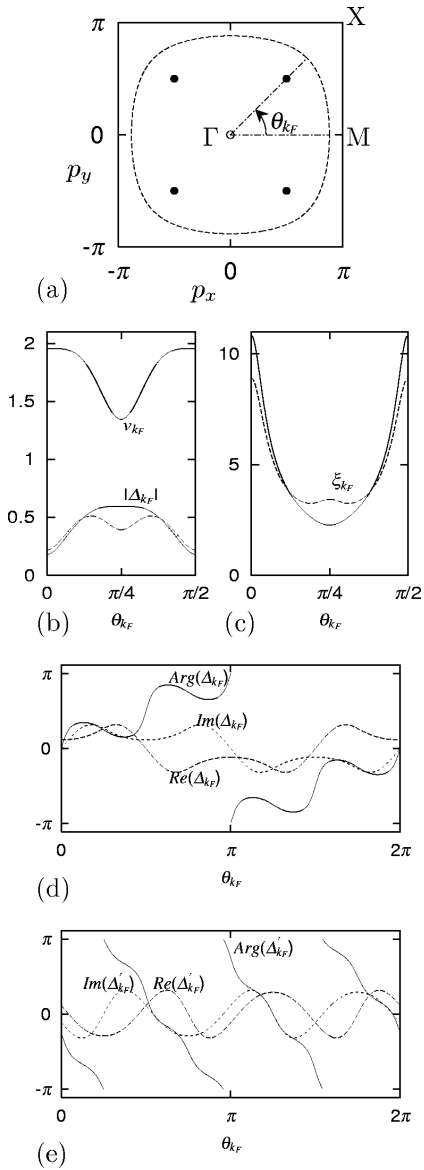}
\end{center}
\caption{
(a) Two dimensional Fermi surface used in our calculation.
We set parameters as  $t'=0.47t$ and $\mu=1.2t$,
which are appropriate to the $\gamma$-sheet in ${\rm Sr_2 Ru O_4}$.
In the phase of the superconducting gap function
$\Delta_+({\bf p}) = \sin{p_x}+{\rm i}\sin{p_y}$ and
$\Delta'_+({\bf p}) = \sin{(p_x+p_y)}+{\rm i}\sin{(-p_x+p_y)}$,
the $\Gamma$ point gives a winding $+1$.
Futher, $\Delta'_+({\bf p})$ has four additional singularities
at $(\pm \pi/2, \pm \pi/2)$(closed circles in the figure),
which give each a winding $-1$.
(b) Fermi velocity $|{\bf v}_F({\bf p})|$(arbitrary unit) and the
amplitude of the gap functions $|\Delta_+({\bf p})|$(solid line) and
$|\Delta'_+({\bf p})|$(dotted line) along the Fermi surface line.
$\theta_{k_{\rm F}}=0$ ($=\pi/4$) corresponds to $\Gamma$-M ($\Gamma$-X)
direction.
(c) The directional dependent coherence length
$\xi({\bf k})=|v_{\rm F}({\bf p})|/|\Delta({\bf p})|$
along the Fermi surface line.
Solid (dotted) line is for the $|\Delta_+({\bf p})|$
($|\Delta'_+({\bf p})|$ ) case.
(d) The phase of $\Delta_+({\bf p})$ along the  Fermi surface line.
We also show ${\rm Re}\Delta_+({\bf p})$ and ${\rm Im}\Delta_+({\bf p})$.
(e) The phase of $\Delta'_+({\bf p})$ along the  Fermi surface line.
}
\label{fig:FS}
\end{figure}

Furthermore we consider the case of NNN pairing interaction, in
contrast to NN pairing mentioned above. The gap function then
corresponds to the form $\sin{(p_x +p_y)}\pm{\rm i}\sin{(-p_x +p_y)}$.
We will see that this case gives $\mp 3$ winding of the Cooper pair
on our considering Fermi surface and belongs therefore to a different
topological class. In our study we will see the role of the
Cooper pair winding.

After describing our formulation of the BdG theory on tight-binding model
in Sec. \ref{sec:Formulation},
we explain the pair potential structure on the Fermi surface
in  Sec. \ref{sec:PPFS}.
We consider the pair potential structure of the vortex lattice state
in Sec. \ref{sec:PairPotential},
and the electronic state in   Sec. \ref{sec:LDOS},
comparing P-vortex and N-vortex both in the NN site pairing case
and the NNN site pairing case.
The last section is devoted to the summary and discussion of our results.

\section{Bogoliubov-de Gennes theory on tight-binding model}
\label{sec:Formulation}

We begin with the tight-binding model on a two-dimensional
square lattice of Ru-site, and introduce a pairing interaction
between NN sites or NNN sites.
The Hamiltonian for the system in a magnetic field is given by
\begin{eqnarray}
{\cal H}&=&
-\sum_{i,j,\alpha} \tilde{t}_{i,j}a^{\dagger}_{i,\alpha} a_{j,\alpha}
+\tilde{V}_{\rm pair}
\label{eq:H1}
\end{eqnarray}
with the creation (an annihilation) operator $a^{\dagger}_{i,\alpha}$
($a_{i,\alpha}$) for spin $\alpha=\uparrow$, $\downarrow$ and
site $i=(i_x,i_y)$.
The transfer integral is expressed as
\begin{equation}
\tilde{t}_{i,j}=t_{i,j} \exp [ {\rm i}\frac{\pi}{\phi_0}\int_{{\bf
r}_i}^{{\bf r}_j} {\bf A}({\bf r}) \cdot {\rm d}{\bf r} ]
\label{eq:Ht}
\end{equation}
with the vector potential
${\bf A}({\bf r})=\frac{1}{2}{\bf H}\times{\bf r}$ in the symmetric gauge,
and the flux quantum $\phi_0$.
For NN-hopping, $t_{i,j}=t$ and for NNN-hopping in diagonal direction
of the square lattice plaquette, $t_{i,j}=t'$.
To reproduce the Fermi surface topology of the $\gamma$-sheet in
${\rm Sr_2RuO_4}$, we set $t'=0.47t$,\cite{Ng} in order to obtain a
closed electron-like Fermi surface for a filling of $ n \approx 1.12 $,
i.e., beyond half filling. Keeping this condition, our results do not vary
qualitatively with different parameters.

The spin part of pairing interaction between $i$- and $j$-sites is decomposed
to the singlet component $g_{0,ji}$ and triplet components
$g_{x,ji}$, $g_{y,ji}$, $g_{z,ji}$ as follows,
\begin{eqnarray}
\tilde{V}_{\rm pair}=&&
\frac{1}{2}\sum_{i,j,\alpha_1 \sim \alpha_4}\sum_{m=0,x,y,z}
g_{m,ji}(\sigma_m{\rm i}\sigma_y)^\dagger_{\alpha_3 \alpha_1}
(\sigma_m{\rm i}\sigma_y)_{\alpha_2 \alpha_4}
\nonumber \\ && \times
(a_{i \alpha_1} a_{j \alpha_3})^\dagger a_{i \alpha_2} a_{j \alpha_4}
\label{eq:HV}
\end{eqnarray}
with the Pauli matrices $\sigma_x$, $\sigma_y$, $\sigma_z$
and a unit matrix $\sigma_0$.
The subscripts $\alpha_1 \sim \alpha_4$ are spin indices.
If we consider the following interaction
\begin{eqnarray}
\tilde{V}_{\rm pair}
=\frac{1}{2}\sum_{i,j}( &&
V_{ji} n_j n_i + J_{x,ji} S_{x,j} S_{x,i}
\nonumber \\ &&
+ J_{y,ji} S_{y,j} S_{y,i}+ J_{z,ji} S_{z,j} S_{z,i} )
\label{eq:HV2}
\end{eqnarray}
using the number density operator $n_i$ and the spin density
operator $S_{x,i}$,
$S_{y,i}$, $S_{z,i}$ at each site, the pairing interactions are obtained
as
\begin{eqnarray} &&
2 g_{0,ji}=V_{ji} -( J_{x,ji}+J_{y,ji}+J_{z,ji})/4, \\ &&
2 g_{x,ji}=V_{ji} +(-J_{x,ji}+J_{y,ji}+J_{z,ji})/4, \\ &&
2 g_{y,ji}=V_{ji} +( J_{x,ji}-J_{y,ji}+J_{z,ji})/4, \\ &&
2 g_{z,ji}=V_{ji} +( J_{x,ji}+J_{y,ji}-J_{z,ji})/4.
\label{eq:HV3}
\end{eqnarray}

The superconducting order parameter is decomposed to
\begin{eqnarray} &&
\Delta_{ji, \alpha_2 \alpha_4}=\sum_{m=0,x,y,z}
d_{m,ji}(\sigma_m{\rm i}\sigma_y)_{\alpha_2 \alpha_4}  .
\label{eq:op}
\end{eqnarray}
The self-consistent condition for the $d$-vector is written as
\begin{eqnarray} &&
d_{m,ji}=g_{m,ji} \sum_{\alpha_1 \alpha_3}
(\sigma_m{\rm i}\sigma_y)^\dagger_{\alpha_3 \alpha_1}
\langle a_{i \alpha_1} a_{j \alpha_3} \rangle.
\label{eq:SCd}
\end{eqnarray}
In this paper, we set $g_{0,ji}=g_{x,ji}=g_{y,ji}=0$ so that
$d_{0,ji}=d_{x,ji}=d_{y,ji}=0$, since we consider the case of
the spin triplet pairing with $d_z$ only.
Thus, when we assume the pairing interaction works only between NN
(NNN) sites, we set $g_{z,ji}=g_{z}$ for NN (NNN) site pairs,
and otherwise  $g_{z,ji}=0$.

In terms of the eigen-energy $E_\epsilon$ and the wave functions
$u_\epsilon({\bf r}_i)$, $v_\epsilon({\bf r}_i)$ at $i$-site,
the Bogoliubov-de Gennes equation is given by
\begin{equation}
\sum_i
\left( \begin{array}{cc}
K_{ji} & d_{z,ji} \\ d^\dagger_{z,ji} & -K^\ast_{ji}
\end{array} \right)
\left( \begin{array}{c} u_\epsilon({\bf r}_i) \\ v_\epsilon({\bf r}_i)
\end{array}\right)
=E_\epsilon
\left( \begin{array}{c} u_\epsilon({\bf r}_j) \\ v_\epsilon({\bf r}_j)
\end{array}\right) ,
\label{eq:BdG1}
\end{equation}
where
$K_{\sigma,i,j}=-\tilde{t}_{i,j} -\mu \delta_{i,j} $
and $\epsilon$ is an index of the
eigenstates.\cite{Wang,Takigawa,TakigawaJPSJ}

The self-consistent condition in Eq. (\ref{eq:SCd}) is reduced to
\begin{eqnarray} &&
d_{z,ji}=g_{z,ji} ( \langle a_{j\downarrow} a_{i\uparrow} \rangle
+  \langle a_{j\uparrow} a_{i\downarrow} \rangle  )
\label{eq:SCdz}
\end{eqnarray}
with
\begin{eqnarray} &&
\langle a_{j\downarrow} a_{i\uparrow} \rangle
=\sum_\epsilon v^\ast_{\epsilon}({\bf r}_j)
               u_{\epsilon}({\bf r}_i) f(E_\epsilon),
\label{eq:SCdza1}
\\ &&
\langle a_{j\uparrow} a_{i\downarrow} \rangle
=\sum_\epsilon u_{\epsilon}({\bf r}_j)
              v^\ast_{\epsilon}({\bf r}_i)  f(-E_\epsilon),
\label{eq:SCdza2}
\end{eqnarray}
and the Fermi distribution function $f(E)$.
The spin-triplet pair potential satisfies the relation
$d_{z,ij}=-d_{z,ji}$, i.e. it corresponds to odd-parity pairing.

The orbital part of the pair potential at each site $i$ can be
decomposed into a $\sin p_x$ and a $\sin p_y$ component as
\begin{eqnarray} &&
\Delta_{p_x}({\bf r}_i) =
( \Delta_{ \hat{x},i}
 -\Delta_{-\hat{x},i} )/2,
\label{eq:dpx}
\\ &&
\Delta_{p_y}({\bf r}_i) =
( \Delta_{ \hat{y},i}
 -\Delta_{-\hat{y},i} )/2
\label{eq:dpy}
\end{eqnarray}
in the NN site pairing case, or into a
$\sin{(p_x+p_y)}$ and a $\sin{(-p_x+p_y)}$ component as
\begin{eqnarray} &&
\Delta_{p_x+p_y}({\bf r}_i) =
( \Delta_{ \hat{x}+\hat{y},i}
 -\Delta_{-\hat{x}-\hat{y},i} )/2,
\label{eq:dpx+py}
\\ &&
\Delta_{-p_x+p_y}({\bf r}_i) =
( \Delta_{-\hat{x}+\hat{y},i}
 -\Delta_{ \hat{x}-\hat{y},i} )/2
\label{eq:dpx-dpy}
\end{eqnarray}
in the NNN site pairing case.
In Eqs. (\ref{eq:dpx})-(\ref{eq:dpx-dpy}) we denoted
\begin{equation}
\Delta_{\hat{e},i}=d_{z,i,i + \hat{e}}
\exp[{\rm i}\frac{\pi}{\phi_0}
\int_{{\bf r}_i}^{({\bf r}_i+{\bf r}_{i + \hat{e}})/2}
{\bf A}({\bf r}) \cdot {\rm d}{\bf r}].
\label{eq:dOP2}
\end{equation}
When the pair potential is uniform, our BdG formulation is
reduced to the conventional theory for $p$-wave superconductors
with the pairing functions $\sin p_x$ and $\sin p_y$, or
$\sin{(p_x+p_y)}$ and $\sin{(-p_x+p_y)}$, by the Fourier
transformation to the momentum space.
For $\sin p_x \pm {\rm i}\sin p_y$-wave superconductivity, we define the
pair potential as $\Delta_{\pm}({\bf r}_i) \equiv \Delta_{p_x}({\bf r}_i)
\mp{\rm i}\Delta_{p_y}({\bf r}_i) $, and
for $\sin{(p_x+p_y)} \pm {\rm i}\sin{(-p_x+p_y)}$-wave superconductivity,
$\Delta'_{\pm}({\bf r}_i) \equiv \Delta_{p_x+p_y}({\bf r}_i)
\mp{\rm i}\Delta_{-p_x+p_y}({\bf r}_i) $.

To investigate the electronic structure around the vortex,
we calculate the LDOS
$N(E,{\bf r}_i)=N_{\uparrow}(E,{\bf r}_i)+N_{\downarrow}(E,{\bf r}_i)$
at the $i$-site, where
\begin{eqnarray} &&
N_{\uparrow}(E,{\bf r}_i)
=\sum_\epsilon  |u_\epsilon({\bf{r}}_i)|^2\delta(E-E_\epsilon), \\ &&
N_{\downarrow}(E,{\bf r}_i)
=\sum_\epsilon  |v_\epsilon({\bf{r}}_i)|^2\delta(E+E_\epsilon)
\label{eq:LDOS}
\end{eqnarray}
for up-spin and down-spin contributions, respectively.
The electron number density at each site is given
by $n({\bf r}_i)=n_{\uparrow}({\bf r}_i) +n_{\downarrow}({\bf r}_i)$ with
\begin{eqnarray} &&
n_{\uparrow}({\bf r}_i)
=\int{\rm d}E N_{\uparrow}(E,{\bf r}_i)f(E)
=\sum_\epsilon |u_\epsilon({\bf r}_i)|^2 f(E_\epsilon ),
\label{eq:BdGn1} \\&&
n_{\downarrow}({\bf r}_i)
=\int{\rm d}E N_{\downarrow}(E,{\bf r}_i)f(E)
=\sum_\epsilon |v_\epsilon({\bf r}_i)|^2 f(-E_\epsilon ) .
\label{eq:BdGn2}
\end{eqnarray}

We consider a system with a square unit cell of
$N_r \times N_r$ sites, where two vortices are accommodated.
The NN vortices of the square vortex lattice are located at the $45^\circ$
directions from $a$-axis of the atomic structure.
This vortex lattice configuration is suggested from the neutron
scattering experiment.\cite{Riseman}
By introducing the quasi-momentum of the magnetic Bloch state
\begin{equation}
u_\epsilon({\bf r})=\tilde{u}_{{\bf k},\epsilon_{\bf k}}({\bf r})
{\rm e}^{{\rm i} {\bf k}\cdot{\bf r}}, \qquad
v_\epsilon({\bf r})=\tilde{v}_{{\bf k},\epsilon_{\bf k}}({\bf r})
{\rm e}^{{\rm i} {\bf k}\cdot{\bf r}}
\label{eq:uv-function}
\end{equation}
with
\begin{equation}
{\bf k}={2 \pi \over c N_r N_k}(l_x,l_y), \qquad
(l_x,l_y=1,\cdots, N_k)
\label{eq:kpoint}
\end{equation}
we obtain the wave function under the periodic boundary condition
whose region covers $N_k \times N_k$ unit cells
($c$ is a lattice constant).
The eigenstate $ \epsilon $ of Eq. (\ref{eq:BdG1}) can then be
labelled by the quasi-momentum ${\bf k}$ and the eigenenergy
 $\epsilon_{\bf k}$ based on Eq. (\ref{eq:uv-function}).

Our calculation starts from the initial state of a square vortex
lattice solution $\Psi_0({\bf r})$ in the lowest Landau level.
Thus, when the dominant superconductivity is $\Delta_{\pm}$ in the
NN-site pairing case, the initial state is given by
$(\Delta_{p_x}({\bf r}_i) ,\Delta_{p_y}({\bf r}_i) )
=(1,\pm{\rm i}) \Psi_0({\bf r})$.
$\Delta'_{\pm}$ in the NNN site pairing corresponds to
$(\Delta_{p_x+p_y}({\bf r}_i) ,\Delta_{-p_x+p_y}({\bf r}_i) )
=(1,\pm{\rm i}) \Psi_0({\bf r})$.
We iterate the calculation of Eqs. (\ref{eq:BdG1})-(\ref{eq:SCdza2})
until convergence is achieved.
We typically consider the case $N_r=20$ and $g_{z,ji}=-1.0t$.
The spatially averaged electron density is set to
$n =\overline{n({\bf r}_i)} \sim 1.12$ by tuning the chemical potential $\mu$.
We consider a sequence of temperatures between $T=0$ and $ T_c$.
With our choice of parameters, we find $T_{\rm c}\sim 0.27t$.
In the figures of this paper, we mainly show the case for $T=0.1t$.
We consider  two cases for the position of the vortex center.
In one case, the vortex center is located just on the atomic site.
We call it as site-centered vortex. For the
other case, it is located in the middle of square plaquette
surrounded by four atomic sites.
We call it as plaquette-centered vortex.
We mainly focus the site-centered vortex case.
We obtain qualitatively the same pair potential structure and the
electron number density structure in both cases.

\section{pair potential structure on the Fermi surface}
\label{sec:PPFS}

Before discussing the vortex structure, we analyze the shape of the
Fermi velocity and the superconducting gap function on the
Fermi surface at zero field.
The two dimensional Fermi surface of our model is shown in
Fig. \ref{fig:FS}(a), corresponding to the $\gamma$-sheet of ${\rm
  Sr_2RuO_4}$.  Along this Fermi surface line, the angle dependence of
the Fermi velocity $v_{\rm F}=|{\bf v}_{\rm F}|$ is depicted in
Fig. \ref{fig:FS}(b). It has a maximum for $\theta_{k_{\rm F}}=0$
($\Gamma$-M direction) and
minimum at  $\theta_{k_{\rm F}}=\pi/4$ ($\Gamma$-X direction), which
results from the vicinity of a van Hove singularity in the $\Gamma$-M
direction. In the same figure also the magnitude of the gap function
$\Delta_+({\bf p})= \sin p_x +{\rm i} \sin  p_y$ on the Fermi surface
is plotted.
$|\Delta_+({\bf p})|$ has minimum at $\theta_{k_{\rm F}}=0$.
The resulting directional dependent coherence length $\xi(\theta_{k_{\rm F}})
=v_{\rm F}(\theta_{k_{\rm F}}) /|\Delta_+(\theta_{k_{\rm F}})| $ is
shown in Fig. \ref{fig:FS}(c), where $ \xi $ is largest for
$\theta_{k_{\rm F}}=0$, and
the ratio  $\xi(\theta_{k_{\rm F}}=0)/\xi(\theta_{k_{\rm F}}=\pi/4) $
is nearly 5.
The directional dependence of $\xi(\theta_{k_{\rm F}})$ is
important for the qualitative interpretation of the spatial
distribution of low energy quasiparticles
around the vortex, as discussed in Sec. \ref{sec:LDOS}.
For $\Delta'_{+}({\bf p})=\sin{(p_x+p_y)}+{\rm i}\sin{(-p_x+p_y)}$,
the amplitude $|\Delta'_{+}({\bf p})|$ has minimum at $\theta_{k_{\rm F}}=0$.
While it has local minimum at $\theta_{k_{\rm F}}=\pi/4$,
it is larger than $|\Delta'_{+}(\theta_{k_{\rm F}}=0)|$.
Therefore, $\xi(\theta_{k_{\rm F}})$ has maximum at  $\theta_{k_{\rm F}}=0$
also for this $\Delta'_{+}({\bf p})$.
The ratio $\xi(\theta_{k_{\rm F}}=0)/\xi(\theta_{k_{\rm F}}=\pi/4) $
is decreased to $2.7$.

In the chiral $p$-wave superconductor, it is important to consider
the phase winding of the gap function on the Fermi surface.
In Fig. \ref{fig:FS}(d), we present the phase
${\rm arg}\Delta_+(\theta_{k_{\rm F}})$ with
${\rm Re}\Delta_+(\theta_{k_{\rm F}})$ and
${\rm Im}\Delta_+(\theta_{k_{\rm F}})$.
The phase of $\Delta_+(\theta_{k_{\rm F}})$ undergoes the winding of $+1$
($ \times 2 \pi $), when we follow it around the Fermi surface.
On the other hand, the phase of $\Delta'_{+}(\theta_{k_{\rm F}})$, as shown in
Fig. \ref{fig:FS}(e), acquires a winding $-3$ around the Fermi surface.
The origin of this difference lies in the structure of the
singularities in the two gap functions.
The gap function $\Delta_+({\bf p})$ and $\Delta'_+({\bf p})$ vanish at
the $\Gamma$ point, where their phase shows a $+1$ winding.
Furthermore, $\Delta'_+({\bf p})$ has additional zeros at the four points
$(p_x,p_y)=(\pm \pi/2,\pm \pi/2)$ not present in $\Delta_+({\bf p})$.
These additional singularities of the gap function give each a winding $-1$.
The Fermi surface in Fig. \ref{fig:FS}(a) is enclosing all five
singularities of $\Delta'_+({\bf p})$
(a $+1$ winding point at $\Gamma$ and
four $-1$ winding points at $(\pm \pi/2,\pm \pi/2))$.
Hence, the total winding of $\Delta'_+({\bf p})$ is $-3$.
A sufficiently small Fermi surface centered around the $ \Gamma
$-point, that does not enclose these outer singularities would,
consequently, have only a winding $+1$, like $ \Delta_+({\bf p})$.
This difference of the winding structure around the Fermi surface
affects the low energy quasiparticle structure around the vortex core.

\section{pair potential structure in the vortex lattice state}
\label{sec:PairPotential}

\begin{figure}[tbc]
\begin{center}
\leavevmode
\epsfxsize=8.0cm
\epsfbox{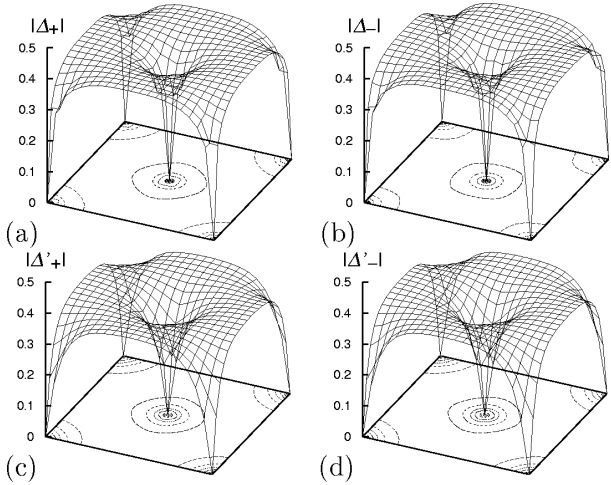}
\end{center}
\caption{
Amplitude of the dominant component of the pair potential
for P-vortex (a) and for N-vortex (b) at $T=0.1t$
for $\sin p_x \pm {\rm i}\sin p_y$-type superconductivity.
We, respectively, plot $|\Delta_{+}({\bf r})|/|g_z|$ and
$|\Delta_{-}({\bf r})|/|g_z|$ in the area of $20 \times 20$ sites, where
vortices are located in the middle and at the four corners of the figure.
(c) and (d), respectively, show the dominant component
$|\Delta'_{+}({\bf r})|/|g_z|$ and $|\Delta'_{-}({\bf r})|/|g_z|$
in the $\sin{(p_x+p_y)}\pm {\rm i}\sin{(-p_x+p_y)}$-wave case.
}
\label{fig:d-main}
\end{figure}

First, we consider the spatial structure of the dominant pair potential
in the vortex lattice state.
Figures \ref{fig:d-main}(a) and \ref{fig:d-main}(b) show
the amplitude of $\Delta_{+}$ for P-vortex
and $\Delta_{-}$ for N-vortex in the case of vortex with winding $+1$,
respectively. Naturally, the amplitude is suppressed around the vortex core.
As seen from contour lines in Fig.  \ref{fig:d-main},
the vortex core has square-like shape,
reflecting the fourfold symmetry
of the Fermi surface structure and the superconducting energy gap
$|\sin p_x \pm{\rm i}\sin p_y|$ in the momentum space.
At low temperature, the orientation of the square shape is different
depending the winding.
In the P-vortex (N-vortex) case,  the corner of the square is directed to
the $0^\circ$- ($45^\circ$-) direction from the $a$-axis, since the
amplitude is more suppressed in this direction.
This difference is smeared with increasing temperature.
At higher temperature near $T_{\rm c}$, the vortex core radius is increased,
and the contour lines both for P- and N-vortices are reduced to the same
structure as the Abrikosov vortex solution $|\Psi_0({\bf r})|$, where
the amplitude is suppressed along the NN vortex direction.

\begin{figure}[tbc]
\begin{center}
\leavevmode
\epsfxsize=8.0cm
\epsfbox{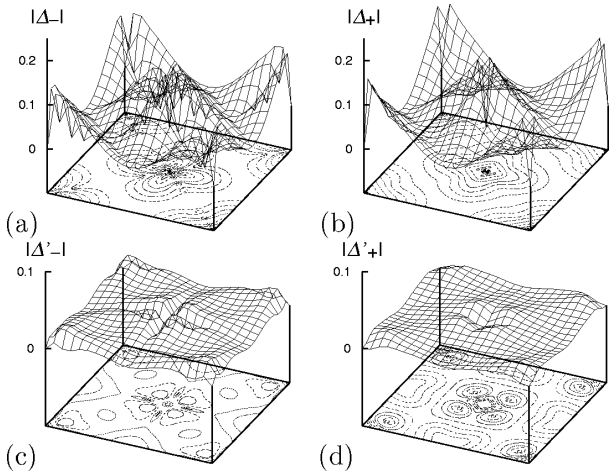}
\end{center}
\caption{
Amplitude of the induced component of the pair potential
for P-vortex (a) and for N-vortex (b)  at $T=0.1t$
for the $\sin p_x \pm {\rm i}\sin p_y$-type state.
We, respectively, plot $|\Delta_{-}({\bf r})|/|g_z|$ and
$|\Delta_{+}({\bf r})|/|g_z|$.
The plotted area is the same as in Fig. \protect{\ref{fig:d-main}}.
(c) and (d), respectively, show the induced component
$|\Delta'_{-}({\bf r})|/|g_z|$ and $|\Delta'_{+}({\bf r})|/|g_z|$
for the $\sin{(p_x+p_y)}\pm{\rm i}\sin{(-p_x+p_y)}$-type state.
}
\label{fig:d-ind}
\end{figure}

Next, we consider the additionally induced component of the pair potential.
In the vortex core region where the dominant component of Fig. \ref{fig:d-main}
shows strongest spatial dependence, the other pairing component is
appears.
It is known that this component occurs also in the framework of
the GL theory, where $\Delta_{+}$ and $\Delta_{-}$ couple to
each other through the gradient terms in the GL equation.\cite{Heeb}
Figure \ref{fig:d-ind} shows the amplitude of $\Delta_{-}$- ($\Delta_{+}$-)
component in the dominant $\Delta_{+}$ ($\Delta_{-}$) pairing channel.
The induced component has the fourfold symmetric structure.
This shape is different between P- and N-vortices.
In the P-vortex case, the induced component $|\Delta_{-}|$
is sharply suppressed along the line of $45^\circ$ direction from $a$-axis.
This difference comes from the winding structure of the induced component.
At higher temperature near $T_{\rm c}$, this difference is smeared.

\begin{figure}[b]
\begin{center}
\leavevmode
\epsfxsize=8.0cm
\epsfbox{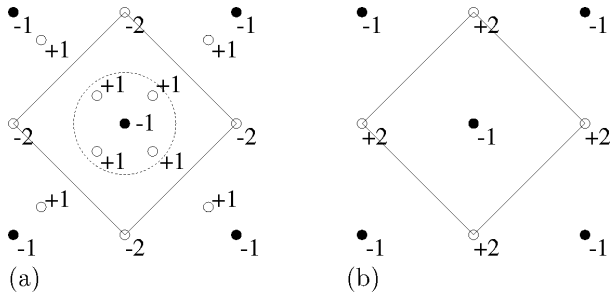}
\end{center}
\caption{
Winding structure of the induced component of the pair potential
in the $\sin p_x \pm {\rm i}\sin p_y$-wave case.
(a) The case of P-vortex at low temperature.
(b) The case of N-vortex.  At higher temperature, P-vortex also has
this configuration.
The plotted area is the same as in Fig. \protect{\ref{fig:d-main}}.
The solid line shows the boundary of the unit cell for the square
vortex lattice.
The vortex center is presented by the solid circles.
The induced component has the winding point at the vortex center and
the points of the open circles.
We also show the winding number at these winding points in the figure.
In (a), total winding number within the dotted circle around the vortex core
is $+3$.
}
\label{fig:winding}
\end{figure}

The winding structure is schematically shown in Fig \ref{fig:winding}.
For the P-vortex, Fig. \ref{fig:winding}(a) shows the basic feature of
winding at low temperature. Fig. \ref{fig:winding}(b) shows the
winding behavior at higher temperature for the P-vortex and for any
temperature for the N-vortex. First, we consider the N-vortex.
The induced $\Delta_{+}$-component of Fig. \ref{fig:d-ind}(b) has winding $-1$
at the vortex center, where the dominant $\Delta_{-}$-component
has winding $+1$.
Since the total of winding number should be $+1$ per vortex both for
the $\Delta_{+}$- and the $\Delta_{-}$-component, the
$\Delta_{+}$-component has also winding $+2$ at the mid points between
NNN vortices (i.e. each corner
of the boundary line of the square vortex lattice unit cell).

Now we consider the P-vortex case at low temperature.
There, the winding structure depends on temperature and
applied magnetic field.
If vortex core radius $r_{\rm core}$ is short enough compared with the
inter-vortex distance $l_{\rm v}=(\phi_0 /H)^{1/2}$ at low temperature,
the induced $\Delta_{-}$-component has winding $-1$ at the vortex center,
and winding $-2$ at the corner of the boundary line.
Furthermore, there are four points with winding $+1$ in the $45^\circ$
direction near the vortex core. Since $|\Delta_{-}|=0$ at these
winding points, the amplitude $|\Delta_{-}|$ is sharply suppressed
along the $45^\circ$
direction, as shown in Fig. \ref{fig:d-ind}(a).
It is known that the induced $s$-wave component
in $d_{x^2-y^2}$-superconductor shows the same type suppression due to
the winding
points.~\cite{Ren1,Ren2,Xu1,Xu2,Berlinsky,FranzKallin,IchiokaInd,IchiokaQCd}
In the limit of an isolated single vortex
($l_{\rm v}/r_{\rm core} \rightarrow \infty$),
four $+1$-winding points approach the $-1$-winding point at the vortex center.
They turn together into a $+3$-winding point at the vortex center.
This structure of winding $+3$ for P-vortex was obtained in the calculation
for an isolated single vortex.\cite{MatsumotoHeeb,Heeb,Kato}
On the other hand, when $r_{\rm core}/l_{\rm v}$ is increased with raising
temperature or applied field, each winding $+1$ point approaches the boundary
line, i.e., the midpoint between NN vortices.
After approaching the boundary, these winding $+1$ points are combined with
winding $-2$ points at the corners of the boundary line and
they become a $+2$-winding point at the corner.
The combination of winding points easily occurs at the boundary,
since the amplitude of the induced component is small there.
As a result, the winding structure is reduced to the same configuration
as that of Fig. {\ref{fig:winding}}(b) at higher temperature.
Therefore, induced components  have the same structure near $T_{\rm c}$
both for P- and N-vortices, while their structures are different
at low temperature.
We have confirmed these winding structures also by solving the
two-component GL theory or the quasiclassical Eilenberger
equation.~\cite{IchiokaU}

The pair potential structure for the $\sin{(p_x+p_y)}\pm{\rm i}\sin{(-p_x+p_y)}$
case is also shown in Figs. \ref{fig:d-main} and \ref{fig:d-ind}.
As for the dominant component in Fig. \ref{fig:d-main},
the difference between $\Delta'_+$ and $\Delta'_-$ is smeared,
as seen from the contour lines.
The amplitude of the induced component in  Fig. \ref{fig:d-ind} is very small,
compared with the $\sin p_x \pm {\rm i}\sin p_y$-wave case.
In this case, the gap amplitude is suppressed along the $0^\circ$
direction on the Fermi surface. Hence, $|\Delta'_-|$ in
Fig. \ref{fig:d-ind}(c) is sharply suppressed along this direction.
In the winding structure of the induced component for
$\Delta'_-$ in Fig. \ref{fig:d-ind}(c),
the $-1$-winding at the vortex center and the $-2$-winding at the corner
is the same as in Fig. {\ref{fig:winding}}(a).
But, the $+1$-winding points around the vortex core is rotated by $45^\circ$
around the vortex center, and is located at the horizontal and vertical
direction at low temperature.
At higher temperatures, the $+1$-winding points approach the boundary of the
unit cell, and the winding structure is reduced to the same as in
Fig. {\ref{fig:winding}}(b).
The transition temperature of the winding structure is shifted to
lower temperatures,
compared with the $\sin p_x \pm {\rm i}\sin p_y$-wave case.
The winding of $\Delta'_+$ in Fig. \ref{fig:d-ind}(d)
has, however, the same structure as Fig. {\ref{fig:winding}}(b).

\begin{figure}[tbc]
\begin{center}
\leavevmode
\epsfxsize=8.0cm
\epsfbox{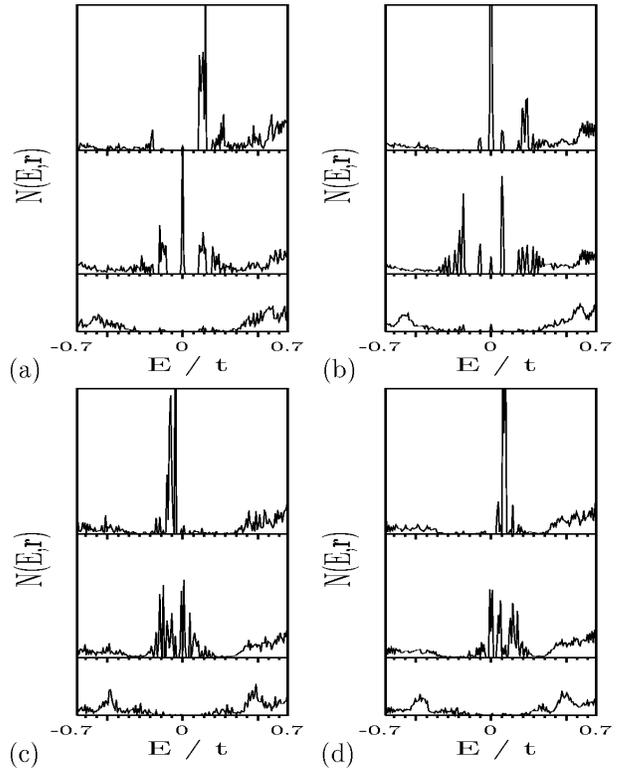}
\end{center}
\caption{
Spectrum of the LDOS $N(E,{\bf r})$ for P-vortex (a) and
for N-vortex (b) in the $\sin p_x \pm {\rm i}\sin p_y$-wave case.
(c) and (d) are, respectively, for
$\Delta'_{+}$ and $\Delta'_{-}$
in the $\sin{(p_x+p_y)}\pm{\rm i}\sin{(-p_x+p_y)}$-type pairing.
The top panel at each figure is for the site just on the vortex center.
The middle panel is for the site next to the vortex center.
The bottom panel is the farthest site from the vortex center, i.e.,
the mid-point between next NN vortices.
}
\label{fig:spectrum}
\end{figure}

\section{electronic state around vortices}
\label{sec:LDOS}

Since the superconducting gap of chiral $p$-wave superconductor opens
a full gap, low energy quasiparticles are bound states around the vortex core.
Therefore, the energy levels are discrete as in $s$-wave superconductors.
However, the energy levels appear at integer points
$E_n \sim n E_\Delta$ ($n$ is an integer) in  chiral $p$-wave
case,~\cite{MatsumotoHeeb,Matsumoto}
while levels appear at half-integer points
$E_n \sim (n+\frac{1}{2}) E_\Delta$  in the $s$-wave
case.\cite{CdGM,HayashiPRL}
Here, $E_\Delta$ is the level spacing of the order $\Delta_0^2 / E_{\rm F}$,
which becomes narrower at higher energy
($\Delta_0$ is a superconducting gap at zero field and
 $E_{\rm F}$ is the Fermi energy).
In Fig. \ref{fig:spectrum}, we show the spectrum of the LDOS at selected
sites inside and outside the vortex core.
There appear some peaks within the superconducting gap.
The energy levels are found to be
$E_1 \sim 0.12t$ for the P-vortex in Fig. \ref{fig:spectrum}(a),
and $E_1=0.08t$ for the N-vortex  in Fig. \ref{fig:spectrum}(b).
We observed also that the peak at $E_0 \sim 0$ vanishes just on the
vortex center for P-vortex.
Instead, sharp peaks appear there at $E_1$ and higher $E$.
The peak at  $E_0$ shoots up sharply at the sites next to the vortex center.
There, the peak at $E_1$  becomes lower, and, moreover,
peaks at $-E_1$ and lower $E$ appear.
For N-vortex, a sharp peak is found at $E_0$ just on the vortex center.
It becomes lower at sites next to the center, and other peaks at
$\pm E_1$, $\pm E_2,\cdots$ emerge in addition.
Outside the vortex core, the spectrum is reduced to that of the zero
field case.
But, small peaks remain at each energy level in this vortex lattice
case due to the low energy quasiparticle transfer between different vortices.
Note also, that the broad gap edge at $E/t=0.3 \sim 0.6$ is due to the
anisotropy of the superconducting gap.

The energy level appearing at the vortex center is consistent with
results of the single vortex calculation.\cite{MatsumotoHeeb,Matsumoto}
For an isolated vortex with circular symmetric structure, the energy
level $E_\epsilon =E_n$ can be labelled by the winding number
of the wave function $u_\epsilon({\bf r})$ around the vortex center.
For the P-vortex and N-vortex, the level $E_n$ corresponds to
$u_\epsilon({\bf r})\sim {\rm e}^{{\rm i}(1-n)\phi}$ and
$u_\epsilon({\bf r})\sim {\rm e}^{{\rm i}n\phi}$, respectively.
On the vortex center, $u_\epsilon({\bf r})$ vanishes except for
the winding $0$ state.
Therefore, the LDOS at the vortex center comes from the  winding $0$ state
of  the wave function.
This corresponds to the level $E_1$ ($E_0$) for the P-vortex (N-vortex).
While $E_0$ state for the P-vortex is a bound state in the vortex core region,
the LDOS vanishes at the vortex center, since the wave function is
given by $u_\epsilon({\bf r})\sim {\rm e}^{{\rm i}\phi}$.
In our calculation for the tight-binding model and the vortex lattice,
the circular symmetry is broken.
Therefore, the wave functions with various winding numbers are mixed for each
energy level.
Even in this case, the characteristics of the energy level
appearing at the vortex center suggested in the single vortex calculation
are qualitatively preserved.
The mixing of the wave functions with different winding numbers may become
larger at higher energy, because the wave functions extend to the boundary
of the vortex lattice unit cell.
Hence, the winding character becomes more smeared for higher energy
state at $n>1$.

Next, we turn to the spectrum for the $\mp 3$-winding case of
$\Delta'_{\pm}$ in the
$\sin{(p_x+p_y)}\pm{\rm i}\sin{(-p_x+p_y)}$-wave superconductivity.
It is presented in Figs. \ref{fig:spectrum}(c) and \ref{fig:spectrum}(d).
At the vortex center, the LDOS has sharp peaks at positive energies for the
$+3$ winding case, shown in \ref{fig:spectrum}(d).
For the $-3$-winding case \ref{fig:spectrum}(c),
peaks appear at the negative energies at the vortex center.
Following the same analysis of Refs. \onlinecite{MatsumotoHeeb} and
\onlinecite{Matsumoto},
we see that the level $E_n$ corresponds to the wave functions
$u_\epsilon({\bf r})\sim {\rm e}^{{\rm i}(2-n)\phi}$ and
$u_\epsilon({\bf r})\sim {\rm e}^{{\rm i}(n+1)\phi}$ for the $+3$ and
$-3$ winding cases, respectively.
Exactly on the vortex center, the LDOS has a peak at $E_2$ for the
$+3$-winding, and at $E_{-1}$ for the $-3$-winding.
This winding character affects the results of Fig. \ref{fig:spectrum}(c)
and (d). The mixing of the energy levels smears this winding
character at higher energy levels.

\begin{figure}[tbc]
\begin{center}
\leavevmode
\epsfxsize=8.0cm
\epsfbox{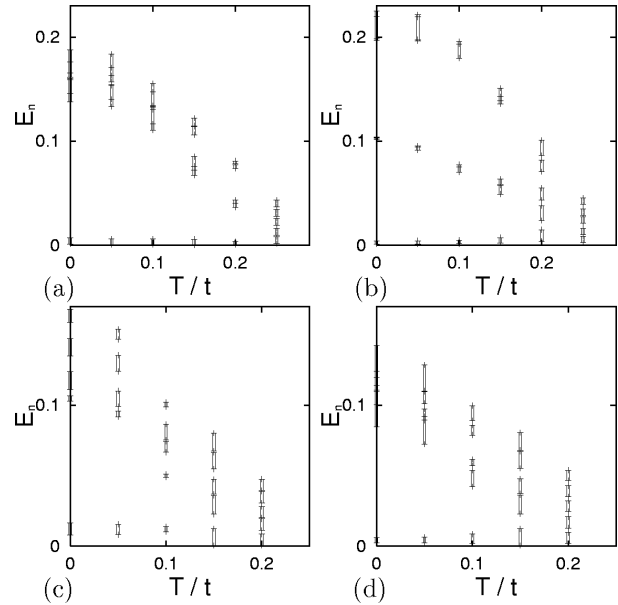}
\end{center}
\caption{
Temperature dependence of the energy levels
for the P-vortex (a) and the N-vortex (b)
in the $\sin p_x \pm {\rm i}\sin p_y$-type state.
(c) and (d) are, respectively, for the
$\Delta'_{+}$ and $\Delta'_{-}$ superconductivity cases
for the $\sin{(p_x+p_y)}\pm{\rm i}\sin{(-p_x+p_y)}$-type pairing.
We show the lower five levels.
Each level has energy width due to the band structure
by the inter-vortex quasiparticle transfer.
}
\label{fig:e-tdep}
\end{figure}

Figure \ref{fig:e-tdep} shows the temperature dependence of the lower
five energy levels.
The quantized energy levels acquire a width by forming band structure due to
the inter-vortex transfer.
Broader band width means larger transfer.
The energy level at $E_0 \sim 0$ splits into two band for $E >0$ and $E<0$.
We show the energy levels of the upper $E_0$ state in
Fig. \ref{fig:e-tdep}. Near $T_{\rm c}$, the levels are distributed
with same level distances, which are small.
Upon lowering of the temperature, the level distance increases.
The level distribution then becomes different depending on the winding
structure, as shown in Figs.  \ref{fig:e-tdep}(a)-(d).
The increasing level distance is due to the shrinking of the vortex core
radius on lowering temperature.
In the BdG theory, the vortex core radius shrinks until the order of the
atomic distance in the clean limit due to the Kramer-Pesch
effect.~\cite{Kramer,Pesch}
The level distance becomes larger at low temperature.
Near zero temperature, this shrinking and level shift saturated.
This is consistent with results of the single vortex
calculation.~\cite{HeebD,HayashiPRL}

\begin{figure}[tbc]
\begin{center}
\leavevmode
\epsfxsize=8.0cm
\epsfbox{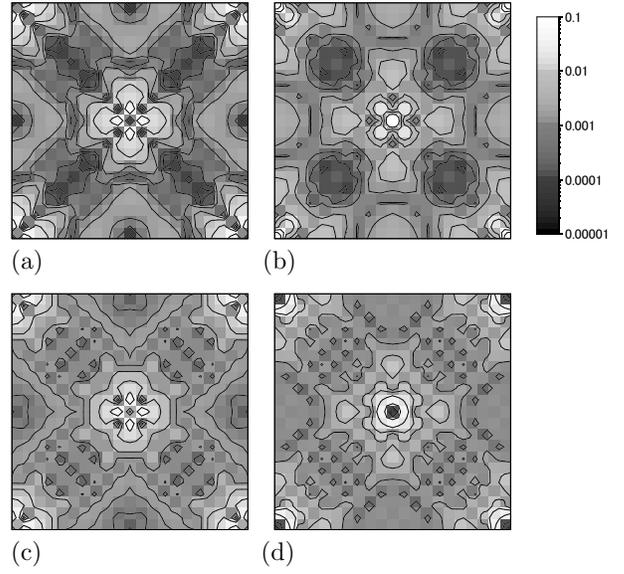}
\end{center}
\caption{
Density plot of the zero-energy LDOS $N(E\sim 0,{\bf r})$
for P-vortex (a) and N-vortex (b)
in the $\sin p_x \pm {\rm i}\sin p_y$-wave case.
(c) and (d) are, respectively, for the
$\Delta'_{+}$ and $\Delta'_{-}$ superconductivity cases
in the $\sin{(p_x+p_y)}\pm{\rm i}\sin{(-p_x+p_y)}$-wave pairing.
The plotted area is the same as in Fig. \protect{\ref{fig:d-main}}.
}
\label{fig:LDOS0}
\end{figure}

Since the low-energy electronic state around the vortex core
determines the low-temperature behavior in the mixed state,
we intend to further investigate the characteristics of the $E_0$ state.
The spatial structure of the LDOS $N(E\sim 0,{\bf r})$ is presented
in Fig. \ref{fig:LDOS0}, where the contribution of $E_0$-states
are integrated. It shows a fourfold symmetric star shape.
Outside the vortex core, $N(E\sim 0,{\bf r})$ extends toward $0^\circ$
direction (i.e., $a$-axis and $b$-axis directions).
In this direction, there is an island of finite LDOS around the midpoints
of NNN vortices.
In the $45^\circ$ direction, the LDOS is suppressed around the mid-point
between NN vortices.
This LDOS structure extending toward the $0^\circ$ direction can be
understood from the view point of the quasiclassical
theory.\cite{HayashiStar,HayashiPRB,Eilenberger}
In this framework, the low energy quasiparticle distribution around the
vortex core is determined by the directional-dependent coherence length
$\xi(\theta_{k_{\rm F}})$ noted in Sec. \ref{sec:PPFS}.
Since $\xi(\theta_{k_{\rm F}}=0)$ is larger than
$\xi(\theta_{k_{\rm F}}=\pi/4)$,
the quasiparticle propagating from the vortex core to $0^\circ$ direction
can extend farther compared with $45^\circ$ direction.
Thus, the LDOS extends toward the $0^\circ$ direction.

At the vortex center, the LDOS vanishes in Fig. \ref{fig:LDOS0}(a) of
the P-vortex case, reflecting the winding structure of the wave function
$u_\epsilon({\bf r})\sim {\rm e}^{{\rm i}(1-n)\phi}$ with $n=0$.
The LDOS is suppressed along the $45^\circ$-direction in the vortex core
region around the vortex center.
In  Fig. \ref{fig:LDOS0}(b) of the N-vortex case, the wave function
$u_\epsilon({\bf r})\sim {\rm e}^{{\rm i}n\phi}$ with $n=0$ does not
vanish at the vortex center, the LDOS has a peak at the vortex center.
It is suppressed along the $0^\circ$-direction around the vortex center
in a small region around the vortex core.

The LDOS for the $\sin{(p_x +p_y)} \pm {\rm i}\sin{(-p_x+p_y)}$-wave case
is shown in Fig. \ref{fig:LDOS0}(c) and (d).
Also in this case, the LDOS extends toward the $0^\circ$ directions
outside of the vortex core.
But, the contrast between the $0^\circ$ direction and the $45^\circ$
direction is smeared in this case.
It comes from the behavior of $\xi(\theta_{k_{\rm F}})$.
For  $\sin{(p_x +p_y)} \pm {\rm i}\sin{(-p_x+p_y)}$-wave case, as shown in
Fig. \ref{fig:FS}(c),
the ratio $\xi(\theta_{k_{\rm F}}=0)/\xi(\theta_{k_{\rm F}}=\pi/4) $
is small compared with the $\sin p_x \pm {\rm i}\sin p_y $-wave case.
Moreover, $\xi(\theta_{k_{\rm F}})$ shows a flat behavior in
a wider range of $\theta_{k_{\rm F}}$ near minimum point $\pi/4$.

In Fig. \ref{fig:LDOS0}(c) of the $-3$-winding case
(in Fig. \ref{fig:LDOS0}(d) of the $+3$-winding case),
The LDOS vanishes at the vortex center, reflecting the winding structure
of the wave function
$u_\epsilon({\bf r})\sim {\rm e}^{{\rm i}(n+1)\phi}$
($u_\epsilon({\bf r})\sim {\rm e}^{{\rm i}(2-n)\phi}$) with $n=0$.
Since the wave function ${\rm e}^{2{\rm i}\phi}$ needs a longer
distance to recover of the order $r^2$ around the singular winding
point compared with
the order $r$ of the wave function ${\rm e}^{{\rm i}\phi}$,
the region of suppression around the vortex core is wider in the
$+3$-winding case. For the $-3$-winding case, the LDOS in the vortex
core region is suppressed along the $45^\circ$ direction.

These LDOS structures are for the site-centered vortex case.
For the plaquette-centered vortex, the LDOS structure is changed within the
vortex core region.
It is due to the trapping effect of the vortex within the  plaquette.
The peak at $E_0$ for P-vortex does not vanish near the vortex center,
since the atomic site is not located exactly on the vortex center.
In the both cases of the vortex center, we obtain qualitatively the same
LDOS structure outside the vortex core.
\begin{figure}[tbc]
\begin{center}
\leavevmode
\epsfxsize=8.0cm
\epsfbox{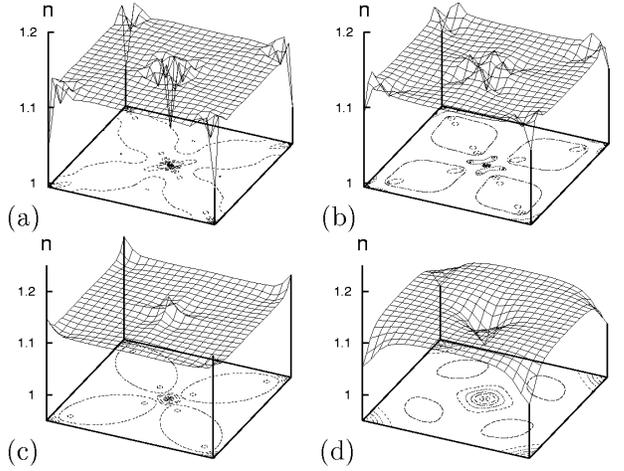}
\end{center}
\caption{
Electron number density $n({\bf r})$ around the vortex
at $T=0.1t$ for P-vortex (a) and N-vortex (b)
in the $\sin p_x \pm {\rm i}\sin p_y$-wave case.
(c) and (d) are, respectively, for the
$\Delta'_{+}$ and $\Delta'_{-}$ superconductivity cases
for $\sin{(p_x+p_y)}\pm{\rm i}\sin{(-p_x+p_y)}$-type pairing.
The plotted area is the same as in Fig. \protect{\ref{fig:d-main}}.
}
\label{fig:n}
\end{figure}

Figure \ref{fig:n} shows the electron number density $n({\bf r})$
around the vortex.
The screening effect~\cite{MatsumotoHeeb} is not included in our calculation.
The number density is slightly changed at the vortex center,
where superconducting gap is suppressed.
This is the vortex charging
effect.\cite{MatsumotoHeeb,Khomskii,Blatter,HayashiCharge,Kumagai}
Near $T_{\rm c}$, $n({\bf r})$  is slightly suppressed for all winding cases.
With lowering temperature, the structure is more eminent and depends on
the Cooper pair winding.
The suppression is small for N-vortex compared with that for P-vortex.
This is consistent with the results of the single vortex
calculation.~\cite{MatsumotoHeeb}
For the $\sin{(p_x+p_y)} \mp {\rm i}\sin{(-p_x+p_y)}$-wave cases,
$n({\bf r})$ is suppressed (enhanced) for the $+3$ ($-3$)-winding case
of the $\Delta'_{-}$ ($\Delta'_{+}$) state.
This $n({\bf r})$ structure at the vortex center in Fig. \ref{fig:n}
is related to the asymmetry between $E>0$ and $E<0$ in the spectrum
shown in Fig. \ref{fig:spectrum}.
When $n({\bf r})$ is largely suppressed [(a) and (d)],
the  spectrum has sharp peaks  at positive $E$ at the vortex center.
If $n({\bf r})$ is enhanced [(c)], the  spectrum has sharp peaks
at negative $E$ at the vortex center.
When the  spectrum has sharp peaks at $E \sim 0$ at the vortex center,
the suppression of $n({\bf r})$ is small [(b)].
The detailed explanation of this mechanism was given in Refs.
\onlinecite{MatsumotoHeeb} and \onlinecite{HayashiCharge}.

At low temperatures, since $n({\bf r})$ shows the Friedel oscillation in the
length order of the atomic lattice constant, $n({\bf r})$ is enhanced around
the vortex along the $0^\circ$
direction before being reduced at the vortex center.
For the $\sin p_x \mp {\rm i} \sin p_y$-wave case,
$n({\bf r})$ exhibits a stronger suppression along the  $45^\circ$ direction
around the vortex core, compared with the $0^\circ$ direction.
For N-vortex, since the suppression along the $45^\circ$ direction extends
farther outside the vortex core, $n({\bf r})$ is reduced on the
lines connecting NN vortices.
This suppression remains up to higher temperatures, while the
suppression at the vortex core is smeared with increasing temperature.
For P-vortex, $n({\bf r})$ is almost constant outside the vortex.
For the $\sin{(p_x+p_y)}\pm {\rm i}\sin{(-p_x+p_y)}$-wave case, the distribution
of $n({\bf r})$ extends toward the $0^\circ$-direction from the
vortex core region instead of the $45^\circ$-direction.

\section{summary and discussions}
\label{sec:summary}

We have investigated the vortex structure in chiral $p$-wave superconductors
by the BdG theory in an extended tight-binding model.
The $\sin p_x +{\rm i} \sin p_y$-type superconductor and the
$\sin p_x -{\rm i}\sin p_y$-type superconductor have different vortex
structures, which are classified as P-vortex and N-vortex.
Around the vortex core in the chiral superconductor,
the opposite chiral component is induced.
The induced component of the P-vortex case and the N-vortex case show
different winding structure at low temperature.
The winding structure of P-vortex depends on temperature,
and it is reduced to that of the N-vortex case near $T_{\rm c}$.
This difference in winding structure affects on the amplitude distribution
of the induced component and the vortex core shape of the dominant
chiral component pair potential.
In spite of the broken circular symmetry around the vortex core
for our tight-binding model and the vortex lattice,
we obtain qualitatively the similar characteristics of the low energy
bound states
around the vortex core as that of the single vortex calculation in a
circular symmetry.~\cite{MatsumotoHeeb,Matsumoto}
For example, we confirmed that the energy levels appearing at the vortex
center and the vortex charging effect are different between P-vortex and
N-vortex.
In the vortex lattice state, the quantized energy levels of bound states
around the vortex core have an energy width by forming a band due
to the inter-vortex transfer.

We have also investigated
the $\sin{(p_x+p_y)} \pm {\rm i} \sin{(-p_x+p_y)}$ type superconductor
by considering NNN site pairing instead of the NN site pairing.
In this case, the Cooper pair winding becomes $\mp 3$ at the Fermi surface.
This winding structure affects the electronic state around the vortex core,
such as the energy level appearing at the vortex center, and the charging
effect at the vortex core region.

In the approach of the BdG theory, we have to consider the case of a
large superconducting gap compared to the Fermi energy,
because of the system size is restricted in the numerical calculation.
Further, in the clean limit, the vortex core radius shrinks to the
order of the atomic distance on lowering temperature due to the Kramer-Pesch
effect.\cite{Kramer,Pesch}
Thus, the level distance of the quantized quasiparticle around vortex core
becomes larger at lower temperature.
Since we are interested in the exotic characteristics of the quasiparticle
at each quantized level, we calculated this case.
In ${\rm Sr_2 Ru O_4}$, this level distance is small.
Then the contribution of the  quantized level appears at very low temperature.
We are also performing the calculation of the quasiclassical theory,
where the level distance is treated as zero.\cite{IchiokaU}
There, qualitatively the same structure was obtained as the result of this
paper for the induced pair potential structure and the LDOS structure outside
the vortex core.

In the chiral $p$-wave superconductors, it is possible to realized a
domain structure consisting of $p_x + {\rm i} p_y$-wave and
$p_x - {\rm i} p_y$-wave superconducting regions.
Therefore, it is important to clarify the difference of the vortex structure
between P-vortex and N-vortex for the investigation of the mixed state
in  ${\rm Sr_2RuO_4}$.
The spatial structure of the LDOS around the vortex include the information
of the anisotropy in the Fermi surface structure and the superconducting gap.
The observation of the LDOS structure is important to get the
information of the superconducting gap structure and the orbital dependence
of the Fermi surface.
It may be examined by scanning tunneling microscopy
experiments.\cite{HayashiStar,Hess1,Hess2,Hess3,Hess4}

\section*{Acknowledgments}

We would like to thank N. Hayashi for helpful comments and discussions.




\begin{references}

\bibitem{Maeno}
Y. Maeno, H. Hashimoto, K. Yoshida, S. Nishizaki, T. Fujita, J. G. Bednorz,
and F. Lichtenberg, Nature {\bf 372}, 532 (1994).

\bibitem{Rice}
T. M. Rice and M. Sigrist, J. Phys.: Cond. Matter {\bf 7}, L643 (1995).

\bibitem{Ishida}
K. Ishida, H. Mukuda, Y. Kitaoka, K. Asayama, Z. Q. Mao, Y. Mori, and Y. Maeno,
Nature  {\bf 396}, 658 (1998).

\bibitem{Luke}
G. M. Luke, Y. Fudamoto, K. M. Kojima, M. I. Larkin, J. Merrin, B. Nachumi,
Y. J. Uemura, Y. Maeno, Z. Q. Mao, Y. Mori, H. Nakamura, and M. Sigrist:
Nature  {\bf 394}, 558 (1998).

\bibitem{Machida}
K. Machida, M. Ozaki, and T. Ohmi,
J. Phys. Soc. Jpn. {\bf 65}, 3720 (1996).

\bibitem{Sigrist}
M. Sigrist and M. E. Zhitomirsky,
J. Phys. Soc. Jpn. {\bf 65}, 3452 (1996).

\bibitem{Hasegawa}
Y. Hasegawa, K. Machida, and M. Ozaki,
J. Phys. Soc. Jpn. {\bf 69}, 336 (2000).

\bibitem{Graf}
M. J. Graf and A. V. Balatsky,
Phys. Rev. B {\bf 62}, 9697 (2000).

\bibitem{Zhitomirsky}
M. E. Zhitomirsky and T. M. Rice,
cond-mat/0102390.

\bibitem{MatsumotoHeeb}
M. Matsumoto and R. Heeb,
cond-mat/0101155.

\bibitem{Heeb}
R. Heeb and D. F. Agterberg, Phys. Rev. B {\bf 59}, 7076 (1999).

\bibitem{HeebD}
R. Heeb, Doctor thesis, ETH Z\"urich (2000).

\bibitem{Kato}
Y. Kato,
J. Phys. Soc. Jpn. {\bf 69}, 3378 (2000).

\bibitem{KatoHayashi}
N. Hayashi and Y. Kato,
private communication.

\bibitem{Matsumoto}
M. Matsumoto and M. Sigrist,
J. Phys. Soc. Jpn. {\bf 68}, 724 (1999).

\bibitem{Wang}
Y. Wang and A. H. MacDonald,
Phys. Rev. B {\bf 52}, 3876 (1995).

\bibitem{Takigawa}
M. Takigawa, M. Ichioka, and K. Machida,
Phys. Rev. Lett. {\bf 83}, 3057 (1999).

\bibitem{TakigawaJPSJ}
M. Takigawa, M. Ichioka, and K. Machida,
J. Phys. Soc. Jpn. {\bf 69}, 3943 (2000).

\bibitem{Himeda}
A. Himeda, M. Ogata, Y. Tanaka, and S. Kashiwaya,
J. Phys. Soc. Jpn. {\bf 66}, 3367 (1997).

\bibitem{Ogata}
M. Ogata,
Int. J. of Mod. Phys. B {\bf 13}, 3560 (1999).

\bibitem{ODS}
D. F. Agterberg, T. M. Rice, and M. Sigrist, Phys. Rev. Lett. {\bf 78},
3374 (1997).

\bibitem{Mackenzie}
A. P. Mackenzie, S. R. Julian, A. J. Diver, G. J. McMullan, M. P. Ray,
G. G. Lonzarich, Y. Maeno, S. Nishizaki, and T. Fujita,
Phys. Rev. Lett. {\bf 76}, 3786 (1996).

\bibitem{MackenzieJS}
A. P. Mackenzie,
J. Superconductivity {\bf 12}, 543 (1999).

\bibitem{Puchkov}
A. V. Puchkov, Z-X. Shen, T. Kimura, and Y. Tokura,
Phys. Rev. B {\bf 58}, 13322 (1998).

\bibitem{Ng}
K. K. Ng and M. Sigrist,
Europhys. Lett. {\bf 49}, 473 (2000).

\bibitem{Miyake}
K. Miyake and O. Narikiyo,
Phys. Rev. Lett. {\bf 83}, 1423 (1999).

\bibitem{Riseman}
T. M. Riseman, P. G. Kealy, E. M. Forgan, A. P. Mackenzie, L. M. Galvin,
A. W. Tyler, S. L. Lee, C. Ager, D. McK. Paul, C. M. Aegerter, R. Cubitt,
Z. Q. Mao, S. Akima, and Y. Maeno,
Nature {\bf 396}, 242 (1998).

\bibitem{Ren1}
Y. Ren, J. H. Xu, and C. S. Ting,
Phys. Rev. Lett. {\bf 74}, 3680 (1995).

\bibitem{Ren2}
Y. Ren, J. H. Xu, and C. S. Ting,
Phys. Rev. B {\bf 53}, 2249 (1996).

\bibitem{Xu1}
J. H. Xu, Y. Ren, and C. S. Ting,
Phys. Rev. B {\bf 52}, 7663 (1995).

\bibitem{Xu2}
J. H. Xu, Y. Ren, and C. S. Ting,
Phys. Rev. B {\bf 53}, 2991 (1996).

\bibitem{Berlinsky}
A. J. Berlinsky, A. L. Fetter, M. Franz, C. Kallin, and P. I. Soininen,
Phys. Rev. Lett. {\bf 75}, 2200 (1995).

\bibitem{FranzKallin}
M. Franz, C. Kallin, P. I. Soininen, A. J. Berlinsky, and A. L. Fetter,
Phys. Rev. B {\bf 53}, 5795 (1996).

\bibitem{IchiokaInd}
M. Ichioka, N. Enomoto, N. Hayashi, and K. Machida,
Phys. Rev. B {\bf 53}, 2233 (1996).

\bibitem{IchiokaQCd}
M. Ichioka, A. Hasegawa, and K. Machida,
Phys. Rev. B {\bf 59}, 8902 (1999).

\bibitem{IchiokaU}
M. Ichioka, unpublished.

\bibitem{CdGM}
C. Caroli, P.-G. de Gennes, and J. Matricon,
Phys. Lett. {\bf 9}, 307 (1964).

\bibitem{HayashiPRL}
N. Hayashi, T. Isoshima, M. Ichioka, and K. Machida,
Phys. Rev. Lett. {\bf  80}, 2921 (1998).

\bibitem{Kramer}
L. Kramer and W. Pesch,
Z. Phys. {\bf 269}, 59 (1974).

\bibitem{Pesch}
W. Pesch and L. Kramer,
J. Low Temp. Phys. {\bf 15}, 367 (1974).

\bibitem{HayashiStar}
N. Hayashi, M. Ichioka, and K. Machida,
Phys. Rev. Lett. {\bf 77}, 4074  (1996).

\bibitem{HayashiPRB}
N. Hayashi, M. Ichioka, and K. Machida,
Phys. Rev. B {\bf 56} 9052 (1997).

\bibitem{Eilenberger}
G. Eilenberger,
Z. Phys. {\bf 214}, 195 (1968).

\bibitem{Khomskii}
D. I. Khomskii and A. Freimuth,
Phys. Rev. Lett. {\bf 75}, 1384 (1995).

\bibitem{Blatter}
G. Blatter, M. Feigel'man, V. Geshkenbein, A. Larkin, and A. van Otterlo,
Phys. Rev. Lett. {\bf 77}, 566 (1996).

\bibitem{HayashiCharge}
N. Hayashi, M. Ichioka, and K. Machida,
J. Phys. Soc. Jpn. {\bf 67}, 3368 (1998).

\bibitem{Kumagai}
K. Kumagai, K. Nozaki, and Y. Matsuda,
Phys. Rev. B {\bf 63}, 144502 (2001).

\bibitem{Hess1}
H. F. Hess, R. B. Robinson, R. C. Dynes, J. M. Valles, Jr., and J. V. Waszczak,
Phys. Rev. Lett. {\bf 62}, 214 (1989).

\bibitem{Hess2}
H. F. Hess, R. B. Robinson, and J. V. Waszczak,
Phys. Rev. Lett. {\bf 64}, 2711 (1990).

\bibitem{Hess3}
H. F. Hess, R. B. Robinson, and J. V. Waszczak,
Physica B {\bf 169}, 422 (1991).

\bibitem{Hess4}
H. F. Hess, Physica C {\bf 185-189}, 259 (1991).

\end{references}
\end{document}